\documentstyle[12pt]{article}
\begin{document}
\newcommand{\eqn}[1]{eq.(\ref{#1})}
\renewcommand{\section}[1]{\addtocounter{section}{1}
\vspace{5mm} \par \noindent
  {\bf \thesection . #1}\setcounter{subsection}{0}
  \par
   \vspace{2mm} } %was 5mm
\newcommand{\sectionsub}[1]{\addtocounter{section}{1}
\vspace{5mm} \par \noindent
  {\bf \thesection . #1}\setcounter{subsection}{0}\par}
\renewcommand{\subsection}[1]{\addtocounter{subsection}{1}
\vspace{2.5mm}\par\noindent {\em \thesubsection . #1}\par
 \vspace{0.5mm} }
\renewcommand{\thebibliography}[1]{ {\vspace{5mm}\par \noindent{\bf
References}\par \vspace{2mm}}
\list
 {\arabic{enumi}.}{\settowidth\labelwidth{[#1]}\leftmargin\labelwidth
 \advance\leftmargin\labelsep\addtolength{\topsep}{-4em}
 \usecounter{enumi}}
 \def\newblock{\hskip .11em plus .33em minus .07em}
 \sloppy\clubpenalty4000\widowpenalty4000
 \sfcode`\.=1000\relax \setlength{\itemsep}{-0.4em} }
%%%%%%%%%%%%%%%%%%%%%%%%%%%%%%%%
%%% some other definitions
\def\ii{{\rm i}}
\def\IR{\relax{\rm I\kern-.18em R}}
\def\IC{\relax\,\hbox{$\inbar\kern-.3em{\rm C}$}}
\def\inbar{\vrule height1.5ex width.4pt depth0pt}
\def\IN{\relax{\rm I\kern-.18em N}}
\def\IP{\relax{\rm I\kern-.18em P}}
\def\wcp{W\IC\IP}
\font\cmss=cmss10 \font\cmsss=cmss10 at 7pt
\def\ZZ{\relax\ifmmode\mathchoice
{\hbox{\cmss Z\kern-.4em Z}}{\hbox{\cmss Z\kern-.4em Z}}
{\lower.9pt\hbox{\cmsss Z\kern-.4em Z}}
{\lower1.2pt\hbox{\cmsss Z\kern-.4em Z}}\else{\cmss Z\kern-.4em
Z}\fi}
\def\bfone{{\bf 1}}
\def\cF{{\cal F}}\def\cN{{\cal N}}\def\cK{{\cal K}}\def\cL{{\cal L}}
\def\dop{{\rm d}\hskip -1pt}
\def\ee#1{{\rm e}^{#1}}
\def\o#1#2{{#1 \over #2}}
\def\Omemind#1{\Omega^{\scriptscriptstyle -\hskip 2pt #1}}
\def\bdm{\begin{displaymath}}
\def\edm{\end{displaymath}}
%%%%%%%%%%%%%%%%%%%%%%%%%%%%%%%%
%%% Text
\begin{flushright}
%SISSA ??/95/EP \\
{\tt hep-th/9511017}
\end{flushright}
\vspace{3mm}
\begin{center}
{\bf R-SYMMETRY OF HETEROTIC N=2 SUPERGRAVITIES\footnote{To appear in the
proceedings of the conference {\sl ``Gauge theories, applied supersymmetry and
quantum gravity''}, Leuven, July 1995.}}\\
\vspace{0.8cm}
M.~BILL\'O\footnote{Address after November 1995: {\it NORDITA, Blegsdamsveij
17, Copenhagen, Denmark}} \\
{\em SISSA} \\
{\em Via Beirut 2, Trieste, Italy} \\
\end{center}
\centerline{ABSTRACT}
\vspace{- 4 mm}  %\end{center}
\begin{quote}\small
The topological twist of N=2, D=4  matter-coupled supergravities requires
a suitable R-symmetry. This symmetry is realised in the effective
supergravities arising at tree level from certain heterotic
compactifications. The set of instanton equations
(topological gauge-fixings) is thus obtained. The conditions that R-symmetry
should satisfy also when these theories are replaced by their ``exact''
quantum-corrected counterparts are investigated.
\end{quote}
\addtocounter{section}{1}
\par \noindent
  {\bf \thesection . Introduction}
  \par
   \vspace{2mm} %was 5 mm
\noindent
Remarkable progress has been made recently towards the
understanding
of some non-perturbative phenomena, both in quantum field theory
and in string theory. In particular, powerful tools emerged
in N=2 supersymmetric theories.
\par
The {\em exact} low energy effective theory of an N=2 super Yang--Mills (SYM)
system was obtained by Seiberg and Witten \cite{SW1}.
\par
At the string level, very strong evidences have been accumulated
\cite{FHSV}-\cite{Antoniadisnew} in favour
of the so-called ``heterotic-type II duality''. To specific heterotic
compactifications, having as
effective 4D field theories N=2 supergravities coupled to $r+1$ vector
multiplets and $m$ hypermultiplets, string-string
duality associates type IIA compactifications on a
Calabi--Yau (CY) manifold with the same spectrum. The Hodge numbers of the
CY must be $h^{1,1} = r + 1, h^{2,1} = m - 1$.
\par
In the heterotic compactifications the dilaton (whose expectation value is
related to the string coupling constant)
sits in a vector multiplet; it belongs instead
to a hypermultiplet in the type II case.
It follows then \cite{DWKLL,stromco} from an N=2 non-renormalization theorem
%that is conjectured to hold beyond perturbation theory \cite{stromco},
%forbids neutral scalar couplings between hyper- and vector multiplets.
%As a consequence of the above two facts,
that the tree level effective theory
of the vector multiplets is exact in the type II compactifications, while it
undergoes perturbative and non-perturbative corrections in the heterotic
case; the opposite happens for the hypermultiplets.
%The supposed duality
%allows thus in principle an exact description of the full low energy
%effective field theory corresponding to a given dual pair of heterotic and
%type II compactifications.
\par
In particular a tree level heterotic effective
theory for the vectors is described in terms of a certain special K\"ahler
manifold $ST(n)$. String corrections modify it; the exact theory
corresponds to a deformed special manifold $\widehat{ST}(n)$. By
heterotic-type II duality $\widehat{ST}(n)$ is identified as the moduli
space of $(1,1)$-forms on the Calabi--Yau manifold representing the type II
compactifying space.
\par
This pattern can be regarded as the analogue, at the level of {\em locally}
supersymmetric theories, of the
evaluation of exact low energy theory for N=2 SYM in terms
of suitable Riemann surfaces \cite{SW1}.
\par
{}From another point of view, N=2 field theories are particularly relevant
because they can be twisted. The topological twist associates them
with
suitable topological field theories (TFT's), providing also a set of
topological gauge-fixings (``instanton equations'') of the latter.
\par
TFT's have often a deep interest in mathematics. For instance, topological
Yang--Mills \cite{wittentop1}
(obtained by twisting N=2 SYM) with ${\rm SU}(2)$ gauge group
is related to Donaldson theory.
The results of Seiberg--Witten on the exact low-energy ${\rm SU}(2)$ SYM
provided, upon twisting, important new tools in Donaldson
theory \cite{topf4d_4}. The ``monopole equation'' that plays
a basic role in this game
is an instance of the instanton equations, obtained by twisting SYM
theories coupled to hypermultiplets \cite{AnFre2,topf4d_8}.
It is therefore very interesting to consider the topological twist
of a generic N=2, D=4 theory, containing supergravity coupled to
hyper- and vector multiplets, and
in particular to consider the structure of the corresponding instanton
equations.
\par
This contribution considers the twist of N=2 effective
supergravities arising from heterotic compactifications, keeping in mind
that the tree level theories are modified by string corrections,
and in certain cases their exact
quantum expressions in terms of CY spaces is conjectured.
\par
In section 2 the basic steps to twist a generic N=2 model are reviewed; between
them, the identification of a R-symmetry. Section 3 recalls the special
geometry of vector multiplets in tree-level heterotic N=2 models. In sec. 4
the requirements that R-symmetry must satisfy are described.
In particular, the R-charges of the dilaton-axion
vector multiplet must differ from the others. By twisting, the set of instanton
equations is obtained in sec. 5. Finally, a case is considered in sec. 6
in which the {\em exact} special geometry is expressed in terms of a
specific CY manifold. The existence of a discrete R-symmetry, allowing
the topological twist, is shown.
\par
\section{Topological twist of N=2, D=4 supergravities}
\noindent
In \cite{wittentop1} Witten derived a topological reinterpretation
of the N=2 YM theory by means of a redefinition of the euclidean  Lorentz
group.
The steps needed to perform this construction in the case of arbitrary
N=2 theories, including also gravity and hypermultiplets, were
derived in \cite{AnFre2,AnFre}, and are as follows:
\begin{enumerate}
\item Systematic use of BRST quantization, prior to the twist
\item Redefinition of the Lorentz group ${\rm SO}(4) = {\rm SU}(2)_L
\times {\rm SU}(2)_R$:
\begin{equation}
\label{leuv1}
{\rm SO}(4)^\prime = {\rm diag} [{\rm SU}(2)_Q \times {\rm SU}(2)_L ]
\times {\rm diag} [{\rm SU}(2)_I \times {\rm SU}(2)_R ] .
\end{equation}
${\rm SU}(2)_I$ is the automorphism group of the N=2 algebra;
${\rm SU}(2)_Q$ acts on the hypermultiplet sector (so that it is
irrelevant in the pure YM case)
\item Redefinition of the BRST charge:
\begin{equation}
Q^\prime_{\rm BRST} = Q_{\rm BRST} + Q^{-0}_{\rm SUSY}
\end{equation}
where $Q^{-0}_{\rm SUSY}$ is  a combination of the components of
the N=2 supersymmetry charges that acquire spin zero after the spin
redefinition (\ref{leuv1}).
\item Redefinition of the ghost numbers of the fields:
\begin{equation}
{\rm g}^\prime = {\rm g} + q_R
\end{equation}
by means of their charges $q_R$ under a suitable R-symmetry, to match the
reinterpretation of the fields as physical
(${\rm g}^\prime = 0$), ghosts (${\rm g}^\prime = 1$), antighosts, \ldots .
\end{enumerate}
In the case of N=2 Yang--Mills, the fields are organized in N=2 vector
multiplets: $(A^I_\mu,\lambda^{IA},\lambda^{I^*}_A,Y^I)$, $\lambda$ being the
gauginos\footnote{The index $A=1,2$ enumerates the supersymmetries;
$\gamma_5\lambda^{IA} = \lambda^{IA}$ and $\gamma_5\lambda^{I^*}_A =
-\lambda^{I^*}_A$ while $\gamma_5\psi_A =\psi_A$ and $\gamma_5\psi^A = -\psi^A$
for the gravitinos}  and $Y$ the gauge scalars.
Their topological reinterpretation is
as follows: $(\mbox{phys., ghost, antigh., gh. for ghost})$.
Accordingly, in the N=2 theory there exists a R-symmetry assigning them
R-charges $(0,1,-1,2)$.
\par
In the general case, besides the vector multiplets,
we have the gravitational multiplet $(V^a_\mu,\psi_A,\psi^A,A^0_\mu)$
including the vierbein $V^a_\mu$, the gravitinos $\psi$ and the graviphoton
$A^0_\mu$, and the hypermultiplets $(q^u,\zeta^\alpha)$, including
the scalars $q^u$, ($u= 1,\ldots 4m$) that span a quaternionic manifold,
and their fermionic partners $\zeta^\alpha$ belonging to the fundamental
representation of ${\rm Sp}(2m)$.
\par
A specific N=2 model is realised by the following geometrical data:
\begin{enumerate}
\item A special K\"ahler manifold for the vector multiplet scalars, of
complex dimension $n$, with $n$ being the number of vector
multiplets.
\item A quaternionic manifold for the hypermultiplet scalars, of quaternionic
dimension $m$, where $m$ is the number of hypermultiplets.
\item A gauge group ${\cal G}\subset {\rm SO}(n)$, acting via special
and quaternionic isometries respectively on the two scalar manifolds.
\end{enumerate}
\par
\section{Vector multiplets in ``heterotic'' N=2 supergravities}
\noindent
The lagrangian and transformation rules of N=2 vector multiplets coupled to
supergravity \cite{skg} are expressed in terms of the ``special geometry''
of the gauge scalar manifold.
The complex gauge scalars $Y^I$ parametrize a K\"ahler
manifold whose K\"ahler potential
\begin{equation}
\label{kalpot}
{\cal K} = -\log \left[ -\ii \Vert \Omega \Vert^2 \right] = - \log \left[ -\ii
({\bar X}^\Lambda F_\Lambda - {\bar F}_\Lambda X^\Lambda) \right]
\end{equation}
is expressed in terms of the $2n+2$-dimensional holomorphic
symplectic section $\Omega
= \left(X^\Lambda(Y), F_\Lambda(Y)\right)$, $\Lambda = 0,I$; it is
invariant under ${\rm Sp}(2n + 2, \IR)$ rotations of $\Omega$.
%As a consequence of (\ref{kalpot}), the Riemann tensor satisfies
%the identity\footnote{The indices $i,j\ldots$ are reserved to generic
%coordinates $z^i$ on the scalar manifold}
%\begin{equation}
%R_{ij^*kl^*}= g_{ij^*} g_{kl^*} + g_{il^*} g_{kj^*} - {\rm e}^{2{\cal K}}
%W_{ikp} {\bar W}_{j^*l^*q^*} g^{pq^*}
%\end{equation}
%where the $W_{ijk}$ are holomorphic and symmetric tensors; ${\rm e}^{\cal K}
%W_{ijk}$ play the role of anomalous magnetic moments in the lagrangian.
\par
The elements of the structural group ${\rm Sp}(2n+2,\IR)$, i.e. matrices
with the block structure
\bdm
\left(\matrix{A & B \cr C & D}\right) , \hskip 1cm
\begin{array}{l} A^T C -C^T A = 0\\ B^T D - D^T B = 0\end{array} ,
\hskip 0.6cm A^T D - C^T B = {\bf 1} ,
\edm
induce coordinate transformations on the scalar manifold while acting, at the
same time, as duality rotations on the symplectic vector of magnetic and
electric field strengths
\bdm
({\cal F}^{-\Lambda}_{\mu\nu}, G^-_{\Lambda\hskip 2pt \mu\nu}) \hskip 0.8cm
\mbox{where} \hskip 0.2cm G^-_{\Lambda\hskip 2pt \mu\nu} = -\ii
{\delta{\cal L} \over \delta {\cal F}^{-\Lambda}_{\mu\nu} } .
\edm
If the scalar manifold admits a continuous or discrete isometry group,
this group must be suitably embedded into ${\rm Sp}(2n + 2,\IR)$:
\begin{equation}
\label{compf}
z \rightarrow \phi(z) \hskip 0.1cm \hookrightarrow \hskip 0.1cm M_\phi \in
{\rm Sp}(2n + 2,\IR) \hskip 0.3cm \mbox{such that} \hskip 0.3cm
\Omega\left(\phi(z)\right) = f(z) M_\phi \cdot \Omega(z) ,
\end{equation}
the holomorphic rescaling $f(z)$ being allowed by equation (\ref{kalpot}).
The corresponding duality transformations on the gauge fields
leave
the system of Bianchi identities plus eq.s of motion form invariant.
They can be of {\em
classical} $(B=C=0)$, {\em perturbative} $(B=0,C\not = 0)$ or {\em
non-perturbative} $(B\not = 0$) type \cite{CDFVP},
according to their projective action
on the coupling matrix ${\cal N}_{\Lambda\Sigma}$, that appears in the
kinetic lagrangian of the gauge fields ${\cal L}\propto {\rm Im}
({\cal N}_{\Lambda\Sigma} {\cal F}^{+\Lambda}_{\mu\nu}
{\cal F}^{+\Sigma}_{\mu\nu})$:
\bdm
{\cal N} \rightarrow (C+ D{\cal N})(A + B{\cal N})^{-1}
\edm
In those N=2 supergravities that arise from heterotic compactifications, the
dilaton-axion scalar field $S = {\cal A} + \ii {\rm e}^D$, with
$\partial_\sigma
{\cal A} = {\epsilon_{\sigma\mu\nu\rho}\over \sqrt{|g|}}
{\rm e}^{2D}\partial^{\mu}B^{\nu\rho}$, belongs to a vector multiplet
$(A^S_\mu,
\lambda^{SA},\lambda^{S^*}_A,S)$, to be adjoined to the ``usual'' vector
multiplets\footnote{We consider therefore now $n+1$ v.m.s. The previous
formulae must be modified in an obvious way, e.g. the structural group is
${\rm Sp}(2n+4,\IR)$, the index $\Lambda$ runs on $0,S,I$, and so on.}.
\par
The existence of a distinguished dilaton-axion direction should have an
intrinsic meaning; in particular we assume that in the twist it is the
dilaton-axion {\em scalar} that remains physical, contrary to the
usual v.m.'s where the physical fields are the vectors.
Twisting the theory with no other vector multiplets apart from that
containing the dilaton
($n=0$), gives a
``string-inspired'' 4D topological gravity. The moduli
space ${\cal M}_\tau$ of a typical instanton configuration (an ALE manifold
with Hirzebruch signature $\tau$) will be $(3+1)\times \tau$ dimensional,
the $+1$ being due
to the axion deformations. Since the observables have typically even ghost
numbers, the selection rule $\sum {\rm g} = {\rm dim} {\cal M}_\tau$ would pose
problems if ${\rm dim}{\cal M}_\tau= 3\tau$, as it would be the case
if only the metric was physical.
\par
The tree-level effective theories that we consider are described by the
choices:
\bdm
\begin{array}{ccc}
\mbox{Special manifold} & \mbox{quatern. manifold} & \mbox{gauge group}\\
\null & \null & \null \\
ST(n)\equiv {{\rm SU}(1,1) \over
{\rm U}(1)}\, \times {{\rm SO}(2,n) \over{\rm SO} (2)\times {\rm SO}(n)}
&
HQ(m )\equiv {{\rm SO}(4,m)\over{\rm SO}(4)\times {\rm SO}(m)} &
{\cal G}\subset {\rm SO}(n)
\end{array}
\edm
This structure  is obtained by certain N=2 truncations of N=4 matter
coupled supergravity, which displays a unique coset structure:
$ {{\rm SU}(1,1)\over {\rm U}(1)}\, \times {{\rm SO}(6,n+m)
\over{\rm SO}(6)\times {\rm SO}(n+m)}$. Different truncations may give
different quaternionic manifolds\footnote{Moreover, for the specific
compactifications for which a type-II dual is proposed \cite{FHSV,Kachru},
the explicit form of the quaternionic manifold that must
have quaternionic dimension equal to the $h^{2,1}$ Hodge number of the dual CY
space,  is not known.}.
Our considerations depend however mainly
on the special manifold $ST(n)$, that is uniquely determined as  it is the only
special manifold
%\cite{}
with explicit factorization of the dilaton \cite{FVP}
(in the ${{\rm SU}(1,1)\over {\rm U}(1)}$ factor).
\par
The special geometry of the $ST(n)$ manifolds is conveniently described
\cite{CDFVP}
in terms of a symplectic section realizing the  following
embedding of the isometry group ${\rm SO}(2,n)\times {\rm SL}(2,\IR)$
into ${\rm Sp}(2n+4,\IR)$:
\def\twomat#1#2#3#4{\left(\begin{array}{cc}#1& #2\\ #3 &
#4\end{array}\right)}
\bdm
A\in {\rm SO}(2,n) \,\hookrightarrow \,\left(\begin{array}{cc}A & {\bf 0}
\\ {\bf 0} & \eta A \eta^{-1} \end{array}\right)\in {\rm Sp}(2n + 4,{\IR})
\edm
\begin{equation} \label{so2nemb}
\left(\begin{array}{cc}a & b\\c & d\end{array}\right) \in {\rm SL}(2,
\IR) \, \hookrightarrow \, \twomat{a\bfone}{b\eta^{-1}}{c\eta}{d\bfone}\in
 {\rm Sp}(2n + 4,\IR)
\end{equation}
where $A^T\eta A=\eta$.
Notice that, in this embedding, the ${\rm SO}(2n)$ transformations
are of classical type. Only ${\rm SU}(1,1)$ generates perturbative and
non-perturbative transformations.
The $S$ field , parametrizing the coset $\o{{\rm SU}(1,
1)}{{\rm U}(1)}$, plays a distinguished role.
The explicit form of the symplectic section,
corresponding to the embedding  of eq. (\ref{so2nemb}), is
\begin{equation}
\label{so2nssec}
(X^{\Lambda}, F_{\Lambda})=(X^{\Lambda}, S \eta_{\Lambda\Sigma}
X^{\Sigma})
\hskip 0.1cm , \hskip 1cm
X^{\Lambda} =
\left(\o 12 (1 + Y^2),\, \o{\ii}{2}(1 - Y^2),\, Y^I\right)
%\begin{array}{c} 1/2\hskip 2pt (1 + Y^2) \\ \ii/2\hskip 2pt (1 -
%Y^2)\\ Y^I\end{array}\right).
\end{equation}
In eq.
(\ref{so2nssec}), $Y^I$ are the Calabi--Visentini coordinates,
for the coset manifold $\o{{\rm SO}(2,n)}{{\rm SO}(2)\times {\rm SO}(n)}$.
The pseudoorthogonal metric $\eta_{\Lambda\Sigma}$
is $(+, +,-, \ldots, -)$.
%\par Notice that, with the choice (\ref{so2nssec}),
% it is not possible to describe $F_\Lambda$ as derivatives of any
%prepotential. \
The K\"ahler potential following from eq. (\ref{kalpot}) is
\bdm
%\label{so2nkpotskn}
\cK=
%\cK_1(S, \bar S)+\cK_2(Y, \bar Y) =
-\log \ii (\bar S - S) - \log {\bar X}^T\eta X .
\edm
The K\"ahler metric has therefore a block diagonal structure; we have
$g_{S\bar S}(S, {\bar S}) = {-1\over (\bar S - S)^2}$ and
$g_{IJ^*}(Y,{\bar Y})= \partial_I\partial_{J^*}\cK$;
At $Y=0$, $g_{IJ^*}$ reduces to $2\delta_{IJ^*}$.
The coupling matrix $\cN$ is such that ${\rm Re}\cN_{\Lambda\Sigma} =
{\cal A}\eta_{\Lambda\Sigma}$; moreover, at $Y=0$, ${\rm Im}\cN_{IJ}$ equals
$\ee D\delta_{IJ}$.
Thus at $Y=0$ the kinetic term for the ordinary gauge vectors
$A^I$ reduces \footnote{We take here  into account the gauge coupling
dependence, via the usual redefinition $A^I \to \o{1}{g} A^I$.} to
 $\o{\rm Im \,S}{g^2}\cF^{I}_{ab} \cF^{I}_{ab}$,
and we can reinterpret
%, as usual in string compactification theory,
$g_{\rm eff} = \o g{\sqrt{{\rm Im} S}}$ as the effective gauge
coupling.
\section{Requirements on R-symmetry}
\noindent
Rigid minimally coupled Yang--Mills theory possesses, as already said, an
R-symmetry; it acts with charge $\pm1$ on the susy parameters, with
charges $(0,1,2)$ on $(A^I_\mu,\lambda^{IA},Y^I)$, commutes with supersymmetry
and is an off-shell symmetry of the action. In the l.e. eff. theory
it is broken to a discrete subgroup by quantum effects \cite{SW1}.
\par
We consider now the requirements that an analogue symmetry  in N=2 local
theories, containing supergravity, must satisfy \cite{rsym}.
\par
We require that the fields of the theory have well defined
charges, so that the R-symmetry group
is either a $U_R(1)$ group if continuous or a cyclic group
$\ZZ_p$ if discrete.
\par
By definition $R$-symmetry acts diagonally with charge $+1$ ($-1$)
on the left-(right)-handed gravitinos (in the same way as
it acts on the supersymmetry parameters in the rigid case):
\begin{equation}
\label{so2n11}
\begin{array}{c}
\psi_A\rightarrow\ee{\ii\vartheta}\psi_A\\
\psi^A \rightarrow\ee{-\ii\vartheta}\psi^A
\end{array}
\hskip 1cm \mbox{i.e.}\hskip 1cm
\begin{array}{l}
q_L(\psi_A) = 1\\
q_R(\psi^A) = -1 .
\end{array}
\end{equation}
$R$-symmetry
must generate isometries\footnote{The indices $i,j,\ldots$ are reserved to
generic coordinates $\{z\}$ on the scalar manifold}
$z^i\rightarrow (R_{2\vartheta}z)^i$
of the scalar metric  $g_{ij^*}$; as such, it has to be embedded into
$Sp(2n+4,\IR)$ by means of  a symplectic matrix:
\begin{equation}
\label{Mmatrix}
 M_{2\vartheta} =
\left(\begin{array}{cc}  a_{2\vartheta} & b_{2\vartheta}
\\c_{2\vartheta} &
d_{2\vartheta}
\end{array}\right)   \, \in \, Sp(2n+4,\IR).
\end{equation}
In general it acts as a {\em R-duality}; only if  $M_{2\vartheta}$
is block-diagonal it corresponds to a symmetry of the lagrangian (we will
see that this happens in the classical case  of the $ST(n)$ manifold).
\par
There is a symplectic action on the section $(X^\Lambda,F_\Lambda)$
induced by $z^i\rightarrow (R_{2\vartheta}z)^i$:
\bdm
\label{fattorello}
(X,F)\rightarrow \ee{2\ii\vartheta}\, {M}_{2\vartheta} \cdot (X,F) .
\edm
Comparing  with eq. (\ref{compf}), we see that the compensating factor $f(z)
_{2\vartheta}$ that is in principle allowed, has to be fixed:
$f(z)_{2\vartheta} = \ee{2\ii\vartheta}$.
Indeed the rescaling factor $f(z)$ in eq. (\ref{compf})
corresponds to a K\"ahler transformation $\ee\cK\rightarrow \ee{\cK + f +
{\bar f}}$; the gravitino $\psi_A$ has K\"ahler weight one, i.e.
$\psi_A\rightarrow \ee{\o f2} \psi_A$.
So $f(z)$ is fixed by eq. (\ref{so2n11}).
This is a very simple but crucial constraint on the
form of the R-symmetry.
\par
The above requirements on the R-symmetry ensure that it is a symmetry of the
bosonic lagrangian, when $M_{2\vartheta}$ is diagonal, or a duality leaving
invariant the set of Bianchi identities and equations of motion.
By explicit analysis
\cite{rsym} of the supersymmetry transformation laws it emerges that such an
R-symmetry does indeed commute with supersymmetry, provided moreover that the
Jacobian matrix $(J_{2\vartheta})^i_j$ of the R-transformation
$z^i \rightarrow (R_{2\vartheta}\,z)^i(z)$ is
{\it covariantly constant}: $\nabla (J_{2\vartheta})^i_j = 0$. This being the
case, we have a symmetry (or a duality) of the full N=2 model.
\par
We add one requirement (which is conceptually independent from the others)
and which pertains to the effective supergravities of heterotic
compactifications. Under the R-action there must be, in the special manifold,
a preferred direction, corresponding to the dilaton--axion multiplet,
whose R-charges are reversed with respect to those of all the other
multiplets.
\par
In the ``classical'' case of the S$T(n)$ models,
the action of  $R$-symmetry is extremely simple. It is nothing else
but the ${\rm SO}(2)\sim {\rm U}(1)$ subgroup in the denominator of the
${\rm SO}(2,n)/{\rm SO}(2)\times {\rm SO}(n)$ coset. The embedding of eq.
(\ref{so2nemb}) induces the transformation:
\bdm
%\label{sknrsim2}
%\begin{array}{l}
\left(\begin{array}{c}X\\F\end{array}\right) \rightarrow \ee{2\ii\vartheta}
\left(\begin{array}{cc}{ m}_{2\vartheta} & 0\\ 0 &
({ m}^T_{2 \vartheta} )^{-1}\end{array}\right)
\left(\begin{array}{c}X\\F\end{array}\right)
%\\ \null \\
\hskip 0.1cm , \hskip 0.2cm
{ m}_{2\vartheta} =
\left(\begin{array}{ccc}\cos 2\vartheta & -\sin 2\vartheta & 0\\
\sin 2\vartheta & \cos 2\vartheta & 0\\
0 & 0 & \bfone_{n\times n}\end{array}\right) ,
%\in SO(2,n) .\end{array}
\edm
where $m_{2\vartheta}$ belongs to ${\rm SO}(2,n)$.
Having chosen the rescaling factor $\ee{2\ii\vartheta}$ as required, the
corresponding transformation on the scalar field is
%\begin{equation}
%\label{sknrsim1}
%\left\{\begin{array}{l} \\ Y^I \rightarrow
%\ee{2\ii\vartheta} Y^I \end{array}\right.
%\hskip 0.5cm \Rightarrow (J_{2\vartheta})^i_j =
%\left(\begin{array}{cc}1 & 0\\ 0 & \ee{2\ii\vartheta}\delta^\alpha_\beta
%\end{array}\right) .
%\end{equation}
$S \rightarrow S$, $Y^I \rightarrow \ee{2\ii\vartheta} Y^I$.
The dilaton-axion is neutral, opposite to the ordinary gauge scalars, as we
wanted. It is easy to check that the Jacobian matrix is covariantly constant.
We need no more checks; this $R$-symmetry is a true symmetry of the
lagrangian and satisfies
all the expected properties.
\par
Notice that in this classical case $b_{2\vartheta}=c_{2\vartheta}=0$,
the matrix (\ref{Mmatrix}) is completely diagonal
and it has the eigenvalues $(\ee{\ii\theta},\ee{-\ii\theta},
1,\ldots,1)$.
\par
In the quantum corrected models ${\widehat ST}(n)$, R-symmetry is a discrete
$\ZZ_p$ group, for some $p\in\IN$.
Consider a coordinate basis $\{ u^{i} \}$ $(i=1,\dots \, n+1)$
that diagonalizes the action of $R_{2\vartheta}$ so that:
\bdm
R_{2\vartheta} \, u^{i} ~=~\alpha^{ q_i} \, u^{i} \quad \quad
q_i = 0,1, \dots , p-1 \, \mbox{mod} \, p\ ;\qquad \alpha^p=1\ .
%\label{drago_1}
\edm
%The $n+1$ integers $q_i$ (defined modulo $p$) are the
%R--symmetry charges of the scalar fields $u_i$.
\label{page7}
A generic $Sp(2n+4,\IR)$ matrix
has eigenvalues\footnote{For convenience, here the index $n+1$
represents the dilaton index previously denoted by $S$}
%\begin{equation}
$\left ( \lambda_0, \ldots , \lambda_{n+1}, {{1}\over{\lambda_0}},
\ldots , {{1}\over{\lambda_{n+1}}} \right )$.
The R--symmetry symplectic matrix $M_{2\vartheta}$ of
eq. (\ref{Mmatrix}), being the generator of a cyclic group
$\ZZ_p$, has eigenvalues
$\lambda_0 = \alpha^{k_0},\ldots ,\lambda_{n+1} = \alpha^{k_{n+1}}$,
where $(k_0,\ldots,k_{n+1} )$ is a new set of
$n+2$ integers defined modulo $p$. These numbers are
the R-charges of the electro--magnetic field strenghts
%\begin{equation}
$\cF_{\mu\nu}^0+{\rm i} \, G_{\mu\nu}^0,$ $\ldots, \cF_{\mu\nu}^{n+1}
+ {\rm i} \, G_{\mu\nu}^{n+1} $.
%\end{equation}
\par
Since what really matters
in the topological twist are the differences of
ghost numbers, the interpretation
of the scalars $u^{i}$ $(i=1, \dots , n)$ as ghost for ghosts and of
the corresponding vector fields as physical gauge fields
requires that
%\begin{equation}
$q_i ~=~k_i \, + \, 2$, for $i=1,\ldots , n$.
%\label{shifto}
%\end{equation}
On the other hand, if the vector partner of the
axion--dilaton field has to be a ghost for ghosts,
the $S$--field itself being physical, we must have
%\begin
$k_{n+1} = q_{n+1} + 2$.
%\label{inversamente}
%\end{equation}
In the last section the existence of such a discrete R-symmetry will be checked
in an explicit example (furnished by heterotic-type II duality) of quantum
$\widehat{ST}(n)$ manifold.
\section{Instanton equations}
\noindent
Having determined a suitable R-symmetry, it is possible to perform all the
steps of section 2 to get the topological theory. In particular, the instanton
conditions are obtained setting to zero the topological BRST variations of the
antighost, at zero unphysical (${\rm g}^\prime \not = 0$) fields.
These variations are obtained from the
N=2 susy transformations of these fields in the untwisted theory.
\par
The antighosts in the gravity-v.m.'s sector are $\psi^A,\lambda^{SA},
\lambda^{I^*}_A$; they have indeed R-charge $-1$ in the N=2 theory. We report
here the form of the instanton equations in the full theory, containing also
the hypermultiplets. To be specific we consider the case in which the h.m.
scalars span the $HQ(m)$ manifold. Being a quaternionic space, three
quaternionic two-forms $\Omega^{-x}$, $x=1,2,3$ exist that are the curvatures
of the one--forms $\omega^{-x}$ gauging the ${\rm SU}(2)_I$ action on $HQ(m)$
\footnote{Another set of one--forms $\omega^{+x}$ gauging the ${\rm SU}(2)_Q$
action, needed for the spin redefinition (\ref{leuv1}), exists. The
$\omega^{\pm x}$ forms  forms can
be explicitly written utilizing a parametrization \cite{rsym}
of $HQ(m)$ that is the quaternionic analogue of the Calabi--Visentini
parametrization of $\o{{\rm SO}(2,n)}{{\rm SO}(2)\times {\rm SO(n)}}$.
The two ${\rm SU}(2)$ groups correspond to the ${\rm SO}(4)\sim {\rm SU}(2)_I
\times {\rm SU}(2)_Q$ decomposition of the ${\rm SO}(4)$ holonomy factor; the
vielbeins have the index structure ${\cal U}^{A\hskip 4pt t}_{\hskip 4pt
{\bar A}}$: $A,{\bar A}, t$ are in the fundamental of ${\rm SU}(2)_I,
{\rm SU}(2)_Q, {\rm SO}(m)$}. The hyperinos appear in the superspace
parametrization of the vielbeins:
${\cal U}^{A~~t}_{~\bar A}$ $=u^{A~~t}_{a\bar A}V^a$ $ +\epsilon^{AB}
\bar \psi_B \zeta^{\bar B  t} \epsilon_{\bar B \bar A}
+ \bar \psi^A \zeta^t_{\bar A}$, and the antighost is $\zeta^{{\bar A}t}$.
\par
Consider also a non-trivial action of the gauge group ${\cal G}$.
Then the K\"ahler and ${\rm SU}(2)
_I$ connections contain a gauge term, and some modifications  occur
in the susy transformation rules (see \cite{dWLVP,DFF,rsym}).
These modifications
are expressed in terms of the Killing vectors and momentum maps
${\cal P}^0_\Lambda$ and ${\cal P}^{-x}_\Lambda$ for the action of $\cal G$ on
the special and quaternionic spaces.
\par
At this point, after Wick rotation\footnote{A non-trivial point is that
together with the Wick rotation, the dilaton-axion field $S = {\cal A} +
\ii\ee{D}$ is rotated to $S= {\cal A} + \ee{D}$}, one considers the susy
variations of the antighosts $\psi^A,\lambda^{SA},\lambda^{I^*}_A,
\zeta^{{\bar A}t}$, makes the index identifications implied by
eq. (\ref{leuv1}), sets to zero the unphysical fields  and projects on suitable
Lorentz components; the details can be found in \cite{rsym}.
The following set of topological gauge-fixing is thus obtained:
\begin{eqnarray}
R^{-ab } \, - \, \sum_{u=1}^3 J_{u}^{-ab}
q^\star \hat \Omemind u   &=& 0\nonumber\\
\partial_a D \, - \, \epsilon_{abcd}\ee D H^{bcd}&=&0 \nonumber\\
{\cal F}^{-\,I\,ab}\, - \, {{g}^2\over{2 \, \exp{D}}} \,
\sum_{u=1}^3 J^{-\, ab}_u {\cal P}_{I}^{-\, u} &=& 0  \nonumber\\
{\cal D}_\mu q^P \, - \, \sum_{u=1}^{3} (j_u)_{\mu}^{~\nu}
\, {\cal D}_\nu q^Q \, ( J_u )_{Q}^{~P}& =& 0 .
\label{istantonequazioni}
\end{eqnarray}
$R^{- ab}$ is the antiselfdual part of the Riemann curvature
two--form ($a,b$ are indices in the tangent of the space time manifold M).
$q^\star \hat \Omemind u$ denotes the pull--back,  via a
gauged--triholomorphic  map:
%\begin{equation}
$q:\,M\longrightarrow HQ(m)$
%\label{triolomorfa}
%\end{equation}
of the  ``gauged" 2--forms $\hat \Omemind u$. $J_u^{-ab}$ is
a basis of anti-selfdual matrices in $\IR^4$.
\par
The first equation (arising from a component of the gravitino variation)
represents the modification of the usual gravitational
instantons ($R^{-ab}=0$) due to the hypermultiplet sector.
\par
The second of equations (\ref{istantonequazioni}), that comes from the
dilatino variation, describes the H--monopole
or {\it axion--dilaton instanton} of \cite{rey}. The H--monopoles have
vanishing stress--energy tensor, so that they do not interfere with
the gravitational instanton conditions. The presence of these H-monopoles is
the main differences with previous analysis of generic 4D TFT's
\cite{AnFre,AnFre2};
as already emphasized, they should allow the calculation of non--vanishing
topological correlators between local observables
as intersection numbers in a moduli--space that has now an overall
complex structure.
\par
The third equation (from the variations of the ordinary gauginos
$\lambda^{I^*}_A$)
expresses the modification (analogue to that of the gravitational instantons)
of Yang--Mills instantons due to the ``hyperinstantons'' \cite{AnFre2}.
\par
The latter are the solutions of the last equation, that arises from the
hyperino variation and  that is the
condition of triholomorphicity of the map $q$,
rewritten with covariant rather than with ordinary derivatives.
\section{R-symmetry in a quantum example}
\noindent
Consider a heterotic N=2, D=4 abelian model with 2 vector multiplets and 87
hypermultiplets.
There exists a CY manifold with the correct Hodge numbers (2,86)
to represent its type II dual, the hypersurface of degree 8 in $\wcp^4_{2,
2,2,1,1}$. Its mirror, that can be described as the vanishing locus of
\begin{equation}
{\cal W}=X_1^8 + X_2^8 + X_3^4 + X_4^4+ X_5^4 -8\psi \, X_1 X_2 X_3 X_4 X_5
-2\phi \, X_1^4 X_2^4 ,
\label{octica}
\end{equation}
has been studied in detail in \cite{candelas}. The moduli space of (2,1)-forms
on this manifold was shown in \cite{BCDFFRSV} (where the analysis was limited
to
the v.m. sector) to be a viable candidate to represent the quantum
$\widehat{ST}(1)$ manifold.
The structure of the chiral ring of ${\cal W}$ ensures that in
the classical limit this moduli space reduces to $ST(1)$.
The hypersurface (\ref{octica}) contains a double covering of the torus
$Z^2+X^4+Y^4-2\phi X^2 Y^2=0$ describing the SW solution of ${\rm SU}(2)$ SYM
theory; such coverings ensure the embedding of the
``rigid'' monodromies at this ``local'' level \cite{BCDFFRSV,KKLMV}. This
heterotic-type II dual pair has recently been thoroughly examined in
\cite{Antoniadisnew}, where also the heterotic perturbative corrections are
reproduced.
\par
The potential (\ref{octica}) admits a $\ZZ_8$ symmetry acting on the moduli as
$\{\psi,\phi\}\rightarrow$ $\{\alpha\psi,-\phi\}$, where $\alpha^8=1$.
This symmetry corresponds to the $\ZZ_4$ discrete R-symmetry of the exact SW
solution; the transmutation into $\ZZ_8$ is due to the double covering.
The ${\rm Sp}(6,\ZZ)$ matrix representing the $\ZZ_8$ action on the
periods\footnote{Recall that, in the type II formulation, the
${\rm Sp}(2n+4,\ZZ)$ group of special geometry acts on the periods
of the holomorphic $(3,0)$ form.}
was derived in \cite{candelas}:
\bdm
Sp(6,\ZZ) \, \ni \, A \, = \, \left (
\matrix{ -1 & 0 & 1 & -2 & 2 & 0 \cr -2 & 1 & 0 & -2 & 4 & 4\cr
0 & 1 & -1 & 0 & 0 & 2 \cr
1 & 0 & 0 & 1 & 0 & 0 \cr
  -1 & 0 & 0 & -1 & 1 & 1 \cr 1 & 0 & 0 & 1 & 0 & -1 \cr  }
\right )
\edm
Its second power $A^2$ represents the $\ZZ_4$ R-symmetry of the exact theory.
Its eigenvalues are $\{-1,\ii,-\ii ; -1,-\ii,\ii\}$. Comparing with the
discussion of pag.~\pageref{page7}, we see that our requirements for discrete
R-symmetry are satisfied; the $\pm\ii$ couple of eigenvalues corresponds
to the graviphoton and gravidilaton directions; the eigenvalue  $-1$
corresponds to the unique
``physical'' vector. Notice moreover that, in the basis in which it is integer
symplectic, the discrete R-symmetry is a non-perturbative duality (the block
$B$ is non-zero); it is not at all the $\ZZ_4$ subgroup of the ``classical''
${\rm U}(1)$ R-symmetry of the $ST(1)$ manifold.
\vspace{4mm}\noindent
{\bf Acknowledgement}: I am extremely grateful to all the co-authors of
\cite{rsym} and \cite{BCDFFRSV} on which this contribution is entirely based,
and particularly P. Fr\'e and A. Van Proeyen for help with this manuscript.
%\vfill\eject

\end{document}